\documentclass[onecolumn,prd,amssymb,amsmath,preprintnumbers,aps,nofootinbib,superscriptaddress,11pt,notitlepage]{revtex4-1}
\usepackage[a4paper, portrait, margin=0.7in]{geometry}

\usepackage[colorlinks=true,urlcolor=blue,linkcolor=blue,citecolor=magenta]{hyperref}

\usepackage{graphicx,slashed,xcolor,hepunits,soul}

\usepackage[english]{babel}

\newcommand{\be}{\begin{equation}}
\newcommand{\ee}{\end{equation}}
\newcommand{\bea}{\begin{eqnarray}}
\newcommand{\eea}{\end{eqnarray}}
\newcommand{\beq}{\begin{eqnarray}}
\newcommand{\eeq}{\end{eqnarray}}

\newcommand{\Eq}[1]{Eq.~(\ref{#1})}
\newcommand{\Eqs}[2]{Eqs.~(\ref{#1}) and (\ref{#2})}

\newcommand{\fgw}{F_{\text{GW}}}

\newcommand{\zetarad}{\zeta_{\text{RD}}}
\newcommand{\delgg}{\delta_\gamma}

\newcommand{\delggp}{\delta_\gamma^{\rm prim}}
\newcommand{\delgw}{\delta_{\text{GW}}}

\newcommand{\delgwp}{\delta^{\rm prim}_{\text{GW}}}
\newcommand{\tgw}{\left. \frac{\Delta T}{T}\right\rvert_{\text{GW}}}
\newcommand{\tcmb}{\left. \frac{\Delta T}{T}\right\rvert_{\text{CMB}}}
\newcommand{\sgw}{S_{\text{GW}}}

\def\mysection#1{{\bf #1.} }

\def\lsim{\mathrel{\rlap{\lower4pt\hbox{\hskip1pt$\sim$}}
     \raise1pt\hbox{$<$}}}         %less than or approx. symbol
\def\gsim{\mathrel{\rlap{\lower4pt\hbox{\hskip1pt$\sim$}}
     \raise1pt\hbox{$>$}}}         %greater than or approx. symbol

\begin{document}

%\widetext
%\leftline{PP-017-28} 

\title{\boldmath \bf \Large Non-Gaussian Stochastic Gravitational Waves from Phase Transitions}

\author{Soubhik Kumar}
\email{soubhik@berkeley.edu}
\thanks{\scriptsize \!\! \href{https://orcid.org/0000-0001-6924-3375}{0000-0001-6924-3375}}
\affiliation{Berkeley Center for Theoretical Physics, Department of Physics, University of California, Berkeley, CA 94720, USA}
\affiliation{Theoretical Physics Group, Lawrence Berkeley National Laboratory, Berkeley, CA 94720, USA
}
\author{Raman Sundrum}
\email{raman@umd.edu}
\affiliation{
Maryland Center for Fundamental Physics, Department of Physics,
University of Maryland, College Park, MD 20742, USA}
\author{Yuhsin  Tsai}
\email{ytsai3@nd.edu}
\affiliation{Department of Physics, University of Notre Dame, IN 46556, USA}

\begin{abstract}
Cosmological phase transitions in the primordial universe can produce anisotropic stochastic gravitational
wave backgrounds (GWB), similar to the cosmic microwave background (CMB). For adiabatic perturbations, the fluctuations in GWB follow those in the CMB, but if primordial fluctuations carry an isocurvature component, this need no longer be true. It is shown that in non-minimal inflationary and reheating settings, primordial isocurvature can survive in GWB and exhibit significant non-Gaussianity (NG) in contrast to the CMB,
while obeying current observational bounds.  While probing such NG GWB is at best a marginal possibility at LISA, there is much greater scope at
future proposed detectors such as DECIGO and BBO. It is even possible that the first observations of inflation-era NG could be made with gravitational wave detectors as opposed to the CMB or Large-Scale Structure surveys.

%satisfy the observational constraints. As we show, even when considering the next generation Large Scale Structure survey and CMB-S4 measurements, future gravitational wave experiments can be the first place to see the non-Gaussian feature of the primordial quantum fluctuations. \YT{mention the model example we give?}
\end{abstract}

\preprint{
}
\maketitle

\tableofcontents
\section{Introduction}
Since the first gravitational wave (GW) signal was detected in 2015~\cite{Abbott:2016blz}, LIGO and VIRGO have continuously observed new astronomical events like the merging of massive black holes and neutron stars~\cite{TheLIGOScientific:2017qsa} with a surprisingly high event rate. Looking forward, we can hope to see \emph{cosmological} signatures at the next generation GW detectors including LISA~\cite{2017arXiv170200786A}, KAGRA~\cite{Somiya:2011np}, LIGO-India~\cite{Unnikrishnan:2013qwa}, Einstein Telescope (ET)~\cite{Punturo:2010zz}, Cosmic Explorer (CE)~\cite{Reitze:2019iox}, DECIGO~\cite{Seto:2001qf,Kawamura:2011zz}, TianGO~\cite{Kuns:2019upi}, BBO~\cite{Crowder:2005nr,Harry:2006fi}, NANOGrav~\cite{McLaughlin:2013ira}, EPTA~\cite{Kramer_2013}, PPTA~\cite{2013PASA3017M} and SKA~\cite{Janssen:2014dka} through the observation of a stochastic GW background (GWB)~\cite{Caprini:2018mtu,Christensen:2018iqi}. 

Such a stochastic GWB can originate, for example, if a new beyond Standard Model (BSM) physics process violently changes the energy density of the early universe through a strong first order phase transition (PT). For a review see~\cite{Mazumdar:2018dfl}. It is possible that the associated BSM particles are too heavy or too weakly coupled to the Standard Model (SM) particles to be studied at particle colliders, in which case the GW frequency spectrum would give us unique insight into the new physics (for example, see~\cite{Schwaller:2015tja,Jaeckel:2016jlh,Croon:2018erz,Cui:2018rwi,Breitbach:2018ddu}). Alternately, the GW signal may be accompanied by BSM collider discoveries, providing invaluable complementary views of the new physics (for example, see~\cite{Arkani-Hamed:2015vfh}). %Even if the associated BSM particles are too heavy or too weakly coupled to the Standard Model (SM) particles to be studied at particle colliders, we can still study their properties using the GW signal they produce through the PT. 
%Different from the photon signals, 
Unlike photons, GW from an early PT can propagate almost freely though the cosmological plasma to reach us. In this way, they can probe deep inside the primordial ``dark age'' before the Big-Bang Nucleosynthesis (BBN).
Therefore, by studying the large-scale anisotropy in the GWB map, we can obtain a complementary insight into the nature of inflationary fluctuations relative to the Cosmic Microwave Background (CMB) and Large-Scale Structure (LSS).

If the PT happens at roughly $T\sim$ TeV, as motivated within many extensions of the SM, the observable universe today would contain $\sim 10^{43}$ Hubble patches during the PT time. Therefore, given the limited angular resolution of GW detectors (e.g., $\delta\theta\sim\mathcal{O}(0.1)$ rad for LISA~\cite{Cutler:1997ta,Kudoh:2004he}), the resulting GW from the PT time  would form a diffuse background since the GW arriving from every direction in the sky is an average of GW originating from a very large number of Hubble volumes undergoing PT. With this large statistics of the number of causally independent Hubble patches that went through the PT, the average temperature when the PT completes
in each finite region of the sky is nearly identical and equals $\sim T_c$, the critical temperature for the PT. %~\footnote{For PT without significant supercooling, nucleation temperature $T_n\approx T_c$, the critical temperature, where the free energies of the two phases becomes equal.}. 
However, since the thermal history of each Hubble patch depends on the primordial energy density fluctuation, each point on the $T=T_c$ surface has a varying distance in redshift from us. This leads to a modulation of the GW energy density $\rho_{\rm GW}(\theta,\phi)$ as a function of angles $\theta,\phi$ on the sky,
\begin{equation}
\rho_{\rm GW}(\theta,\phi)=\bar\rho_{\rm GW}+\delta\rho_{\rm GW}(\theta,\phi)\,,
\end{equation}
and any stochastic GWB produced in the early universe must be anisotropic.\footnote{For ansitropic GW signals from astrophysical sources, see e.g.~\cite{Cutler:2009qv,Cusin:2017fwz,Cusin:2017mjm}.} As discussed in~\cite{Geller:2018mwu}, the GWB power spectrum $\langle\delta\rho_{\rm GW}\delta\rho_{\rm GW}\rangle$ and cross correlation with the CMB $\langle\delta\rho_{\rm GW}\delta\rho_{\rm CMB}\rangle$ can probe the quantum fluctuations of the inflation-era fields that reheated the PT sector, without presuming that this is identical to the source of fluctuations in the CMB and LSS. While such GWB anisotropies would be challenging to measure, it was argued in~\cite{Geller:2018mwu} that they were within the reach of proposed detectors for plausibly strong PT.

Moreover, just as primordial non-Gaussianity (NG) in the CMB or LSS can arise as reflections of interactions of the inflationary (and possibly other) fields, 
and is being actively searched for, 
it is possible for the GWB fluctuations to also exhibit NG. This would be captured by the GWB bispectrum correlator $\langle\delta\rho_{\rm GW}(\hat{n}_1) \delta\rho_{\rm GW}(\hat{n}_2) \delta\rho_{\rm GW}(\hat{n}_3)\rangle$ where $\hat{n}_i$'s for $i=1,2,3$ denote three angles on the sky. Again, this GWB bispectrum does not have to closely approximate that of the CMB if the PT sector and SM are originally reheated by different (combinations of) reheating fields, with different interactions.

In this work, we will study such GWB NG, and the mechanism by which they may be significantly stronger than (the bounds on) NG in the CMB and LSS, and plausibly observable at future GW detectors.
%the fluctuations $\delta\rho_{\rm GW}$ from a cosmological PT can provide invaluable information of how the new particles were populated in the Universe and their interactions. As it is already extremely challenging to examine the thermal history of the universe before BBN, it is even more difficult to get the information of the reheating process that populates the new particles at even earlier times. However, if we get to measure the power spectrum $\langle\delta\rho_{\rm GW}^2\rangle$ of the GWB produced during the PT, we can probe the quantum fluctuations of the source field that reheated the new particles. 
%While such interactions can be very difficult to probe via CMB and LSS, especially in the cases where the sector undergoing PT has a subdominant energy density, we will show below that future GW detectors can still probe a significant part of the relevant parameter space. 
After production, GWs free stream to us and hence the primordial GWB NG would not be significantly affected by gravitational clustering which, however, is important for CMB and (especially) LSS NG.
%Similar to the \emph{cosmological collider} physics that the non-Gaussian (NG) CMB spectrum can come from on-shell interactions between the SM source fields, the observation of the bispectrum $\langle\delta\rho_{\rm GW}^3\rangle$ will indicate the existence of non-trivial interactions between the invisible source fields.

When discussing the three point function of GWB, %some of the existing literature consider the correlation function of the strain $\langle h^3\rangle$. 
it is shown in~\cite{Bartolo:2018evs,Bartolo:2018rku} that the correlation function $\langle h^3\rangle$, with $h$ denoting an individual GW fluctuation, is almost impossible to measure---both due to the propagation of GW in a perturbed background that would de-correlate the phases in $h$, and the difficulty in separating $h$ with nearby frequencies in measurements with finite durations. However the bispectrum $\langle\delta\rho_{\rm GW}^3\rangle$ in the spatial distribution of the GW energy density does not depend on the initial phase of GW and we can always measure this bispectrum after obtaining a map of $\delta\rho_{\rm GW}$ in the GWB. The prospects of GW map-making have been discussed previously in the literature, for a review see e.g.~\cite{Romano:2016dpx}, along with some recent work~\cite{Renzini:2018vkx,Contaldi:2020rht}. For previous discussions on anisotropies and non-Gaussianity of GWB, see~\cite{Bartolo:2019oiq,Bartolo:2019yeu}.

This paper is organised as follows. After reviewing the properties of a stochastic GWB from first order PT, we estimate to what extent the anisotropies of such a background can be visible at future GW detectors. Following this, we discuss NG of GWB both in the case of adiabatic and isocurvature perturbations. We find that for the latter case, provided the astrophysical GW ``foreground'' from binary mergers can be subtracted, future GW detectors would be able to probe primordial NG in a stronger way compared to the CMB and especially, LSS. We then give a simple model that demonstrates that the NG in GW can be significantly large while ensuring theoretical control and obeying observational bounds. Finally, we conclude.

\section{Gravitational waves from phase transitions}
GW from first order PT get generated due to three processes: collisions of nucleating bubble walls, magnetohydrodynamic (MHD) turbulence and sound waves in the plasma
(for recent discussions see~\cite{Caprini:2015zlo,Caprini:2019egz} and references therein).
According to simulations~\cite{Hindmarsh:2015qta,Niksa:2018ofa,Caprini:2019egz}, the MHD turbulence and sound waves can contribute dominantly to the generation of GW compared to bubble collisions, but those are dependent on details of the plasma and model-specifics, and are being actively researched.\footnote{Their relative contributions depend also on efficiency factors that govern how much of the latent heat released during PT goes into accelerating the bubble walls, sound waves and turbulence.}
On the other hand, through the envelope approximation, the contribution of bubble-wall collisions are analytically better understood~\cite{Kosowsky:1991ua,Kosowsky:1992rz,Kosowsky:1992vn,Kamionkowski:1993fg,Huber:2008hg} and therefore, we focus only on this contribution.
%~\footnote{\SKc{As a proof-of-principle, we can consider strongly supercooled PT, in which case, the plasma effects can get significantly diluted such that the dominant signal comes from bubble collisions.}}. 
Given that we will be studying very small GWB and its correlation functions, this is a conservative approach.
Of course, in the context of a given microscopic BSM dynamics responsible for the first order PT, the signals from sound waves and turbulence can dominate, however, that would only increase the GW signals considered below. We note that while the more recently considered ``bulk-flow approximation''~\cite{Jinno:2017fby,Konstandin:2017sat} can change the frequency dependence of GW away from the peak, the peak amplitude roughly matches the one from the envelope approximation for the models considered in~\cite{Lewicki:2020jiv}. Hence we continue to use the result from~\cite{Huber:2008hg} for the peak amplitude, as needed for our discussion on GW anisotropy, using the envelope approximation.
This will allow us to give simple model-independent estimates/lower-bounds of the overall anisotropy and NG in order to see if they can be feasibly detected, while deferring modeling/calculations of the precise frequency spectrum as long as its peak lies within the sensitivity band of proposed detectors. 

The peak energy density of GW from bubble-wall collisions in a thin-wall approximation, expressed in terms of the %energy in CMB photons today ($\rho_{\gamma}$) is
critical density today, is given by \cite{Huber:2008hg}
\begin{equation}
\Omega_{\rm GW}^{\rm peak}h^2 =  1.3\times10^{-6}\left(\frac{H_{\rm PT}}{\beta}\right)^2 \left(\frac{\alpha}{1+\alpha}\right)^2.
\label{eq:rhoGW}
\end{equation}
% \begin{equation}
% \rho_{\rm GW}^{today} = 0.06 \frac{\rho_{PT}^2}{\rho_{total}^2} 
% %\left(H_{PT} \Delta t_{PT} \right)^2 
% \left(\frac{H_{PT}}{\beta}\right)^2\rho_{\gamma},
% \label{eq:rhoGW}
% \end{equation}
%as reviewed in Ref.~\cite{Konstandin:2017sat,Cutting:2018tjt}.
In the above, the parameter $\beta$ essentially captures the inverse duration of the PT, is approximately given by $\beta/H_{\rm PT}\equiv d\ln\Gamma/dt\approx -4+T_n dS_b/dT\rvert_{T_n}$ in terms of the bubble nucleation rate $\Gamma$ and the bounce action $S_b$ at the nucleation temperature $T_n$. The quantity $\alpha=\rho_{\rm vac}/\rho_{\rm rad}$ is the ratio of the vacuum energy density released during the PT to the energy density of radiation bath. %, and the present Hubble parameter is given by $H=100h \text{ km/s/Mpc}$. 
The Hubble parameter at the time of the PT is given by $H_{\rm PT}\approx \sqrt{(8\pi/3)G_N \rho_{\rm tot}}$, and we assume the effective number of degrees of freedom $g_*\sim 100$. Furthermore, we have considered the scenario in which the bubble walls move close to the speed of light, and efficiency factor $\kappa_b\simeq 1$, i.e. most of the latent heat of the PT is used up in accelerating the bubble walls.\footnote{More precisely, $\kappa_b=1-\frac{\alpha_\infty}{\alpha}$ where $\alpha_\infty$ is a threshold value such that if $\alpha>\alpha_\infty$, then wall velocity $v_w\approx 1$. The precise value of $\alpha_\infty$ is model-dependent but can be $\sim 0.01$~\cite{Caprini:2015zlo,Schmitz:2020syl}. With a choice of $\alpha\gtrsim 0.01$, we then get $\kappa_b\simeq 1$.} The quantity $\beta/H_{\rm PT}$ is model-dependent and can range between ${\rm few} - {\cal O}(100)$, in models in the literature with a strong first-order PT~\cite{Randall:2006py,Konstandin:2011dr,Baratella:2018pxi,Megias:2018sxv,Agashe:2019lhy,Agashe:2020lfz}.
The GW frequencies are redshifted and the peak frequency is given by,
\begin{eqnarray}\label{eq:ftoday}
\omega^{\rm peak}_{\rm GW} &=& 0.04\,{\rm mHz}~
\left(\frac{\beta}{H_{\rm PT}}\right) \left(\frac{T_n}{\rm TeV}\right).
\end{eqnarray}

\section{Anisotropic gravitational wave sky}
In the simplest scenario where there is a single, adiabatic source of primordial fluctuations,
such as single-field inflation where the inflaton decays to reheat all sectors, the primordial fluctuations in each sector will be identical.
This means the perturbation of the GWB, follows the almost scale-invariant adiabatic result, and is determined in terms of the primordial power spectrum $A_s$: $\delta\rho_{\rm GW}=\frac{4}{3}\sqrt{A_s}\rho_{\rm GW}\approx 6\times10^{-5}\rho_{\rm GW}$~\cite{Aghanim:2018eyx}, just as for the CMB. 
 Therefore% in terms of $\delta\Omega_{\rm GW}^{\rm peak, ad}\equiv \delta\rho^{\rm ad}_{\rm GW}/\rho_{\rm crit}$,
\begin{eqnarray}
\delta\Omega_{\rm GW}^{\rm peak}h^2&=&\Omega_{\rm GW}^{\rm peak}h^2\times\left(\frac{\delta\rho_{\rm GW}}{\rho_{\rm GW}}\right),\label{eq.delom0}
\\
&\approx& 8\times10^{-11}\left(\frac{H_{\rm PT}}{\beta}\right)^2\left(\frac{\alpha}{1+\alpha}\right)^2.\label{eq.delom}
\end{eqnarray}
%\begin{equation}
%\delta\rho_{\rm GW}^{\rm ad}\approx 2.8\times10^{-6}\frac{\rho_{PT}^2}{\rho_{total}^2} 
%\left(\frac{H_{PT}}{\beta}\right)^2\rho_{\gamma}\,.
%\end{equation}

In analogy to the CMB, we can calculate the two point function of GW perturbation $\delta_{\rm GW}\equiv\delta\rho_{\rm GW}/\rho_{\rm GW}$,
\begin{equation}
C^{\rm GW}(\theta)\equiv\langle\delta_{\rm GW}(\hat{n}_1)\delta_{\rm GW}(\hat{n}_2)\rangle,\,
\end{equation}
and the coefficient of Legendre polynomials $C_{\ell}^{\rm GW}$ from 
\begin{equation}
C^{\rm GW}(\theta)=\frac{1}{4\pi}\sum_{\ell}(2\ell+1)C_{\ell}^{\rm GW}P_{\ell}(\cos\theta)\,,
\end{equation}
with $\hat{n}_1\cdot\hat{n}_2=\cos\theta$.
In the above, $\delta_{\rm GW}$ would get additional imprint from the Sachs-Wolfe (SW) and the integrated Sachs-Wolfe (ISW) effects~\cite{Sachs:1967er,Rees:1968zza}. Since the ISW effect is a known, standard $\Lambda$CDM contribution in our setup, we only consider the SW effect, and take $\delta_{\rm GW}$ as the sum of the primordial density contrast $\delta_{\rm GW}^{\rm prim}$ and Newtonian potential $\Phi$,
$\delta_{\rm GW} = \delta_{\rm GW}^{\rm prim}+4\Phi$. Here we are using the Newtonian gauge,
\begin{equation}\label{eq.new}
ds^2 = a^2(\eta)\left(-(1+2\Phi)d\eta^2+ (1-2\Psi)d\vec{x}^2 \right),
\end{equation}
to characterize metric fluctuations, with $a(\eta)$ being the scale factor written in terms of the conformal time $\eta$ and $\Phi,\Psi$ are Newtonian gauge potentials. More detailed discussions on the characterization of GW anisotropy can be found in e.g.~\cite{Olmez:2011cg,Contaldi:2016koz,Cusin:2017fwz,Cusin:2018rsq,Jenkins:2018nty}. From the CMB observation, besides small deviations coming from the physics of inflation/reheating~\cite{Baumann:2009ds}, we expect the large-scale GWB perturbation should be almost scale-invariant. An exact scale invariant spectrum would correspond to
\begin{equation}\label{eq.Cl}
C_{\ell}^{\rm GW}\propto[\ell(\ell+1)]^{-1}\,. 
\end{equation}

Our ability to observe such GW anisotropies will be limited by detector sensitivity, as well as our understanding and ability to subtract the GW ``foregrounds'' arising from astrophysical mergers. The first of these limitations can be estimated simply as follows.
The best LISA sensitivity, with Signal-to-Noise Ratio (SNR) $>1$, is $\Omega_{\rm GW}^{\rm peak}h^2 \gsim 2\times 10^{-14}$ in the frequency range $\omega_{\rm GW}^{\rm peak}\approx 1-10$ mHz, see e.g.~\cite{Thrane:2013oya,Schmitz:2020syl}. If $\left(\beta/H_{\rm PT}\right)^2=10$ and $T_{PT}\approx10-100$ TeV, the GW frequency is inside this range, and $\delta\Omega_{\rm GW}h^2$ with $\alpha\sim\mathcal{O}(0.1)$ is comparable to the LISA sensitivity for $\ell\lsim$ few. With its expected angular resolution, LISA then may probe perturbations with $\ell\lsim 10$ (see~\cite{Kudoh:2004he,Contaldi:2020rht} for discussions on LISA sensitivity to stochastic GWB). However, due to the asymptotic falloff $\sqrt{C_{\ell}^{\rm GW}}\propto 1/\ell$ $(\ell\gg 1)$ from Eq.~(\ref{eq.Cl}), it is very challenging to probe beyond $\ell\approx \mathcal{O}(10)$ in LISA. This is where proposed space-based missions such as DECIGO~\cite{Seto:2001qf,Kawamura:2011zz} and BBO~\cite{Harry:2006fi,Crowder:2005nr} can make significant progress. The best DECIGO and BBO sensitivity, with SNR $>1$, is around $\Omega_{\rm GW}^{\rm peak}h^2 \gsim 10^{-17}$ (see e.g.~\cite{Corbin:2005ny,Thrane:2013oya}), around a frequency $\omega_{\rm GW}^{\rm peak}\approx 100$ mHz and $T_n\sim 10^4$~TeV.
%The expected sensitivity of BBO is $\approx 10^{-11}\rho_{\gamma}$~[cite] with $\omega_{\rm GW}^{\rm today}\approx 0.1$ mHz. 
Thus we can observe $\delta\Omega_{\rm GW}h^2$ up to $\ell\approx 100$ in DECIGO and BBO even for a shorter PT $\left(\beta/H_{\rm PT}\right)^2=100$ with $\alpha\sim\mathcal{O}(0.1)$. For a recent discussion on angular resolution of future GW detectors for various configurations, including those in deci-Hz range, see~\cite{Baker:2019ync}.

But of course, to achieve the above targets we must be able to identify and ``subtract'' the astrophysical GW foregrounds that can be competitive to the cosmological GW anisotropies in the relevant frequency ranges. 
These can take the form of unresolved or resolvable merger events, such as those involving white dwarfs, neutron stars and black holes (see~\cite{Regimbau:2011rp} for a review). Our ability to extract primordial GWB in the presence of such astrophysical foregrounds has been studied in several papers including~\cite{Cutler:2005qq,Harms:2008xv,Yagi:2011wg,Regimbau:2016ike,Sachdev:2020bkk,Sharma:2020btq,Pieroni:2020rob,Biscoveanu:2020gds,Barish:2020vmy}. In particular, Refs.~\cite{Cutler:2005qq,Harms:2008xv} focus on the subtraction problem for BBO/DECIGO frequencies, and shows that almost all the mergers can be individually subtracted to isolate the cosmological signal~\cite{Cutler:2009qv}. However, all these studies focused on situations which the astrophysical foreground competes with the isotropic cosmological signal. Here, we are considering the possibility that the isotropic cosmological signal is very strong, much larger than the astrophysical foreground, but where the cosmological anisotropies are smaller. Nevertheless, astrophysical foreground subtraction may still be possible in this novel regime, exploiting the cosmological frequency spectrum that can be precisely measured from the large isotropic component. This deserves dedicated study, but we will proceed assuming such subtraction is possible.

\section{Non-Gaussian gravitational waves with adiabatic perturbations}
Before moving on to the GW NG, we first parametrize the CMB NG. Instead of using the conventional harmonic coefficients $a_{lm}$ of the CMB anisotropies, we describe the CMB NG by simply using the CMB density perturbations in momentum space (including the SW gravitational redshift), $\delta_{\gamma}=\delta_{\gamma}^{\rm prim}+4\Phi$. This will be sufficient to illustrate the essential physical effects of adiabatic and isocurvature perturbations on CMB and GW NG. Thus we parametrize the CMB bispectrum as\footnote{also conventionally denoted as the $f_{\rm NL}$ parameter},
\begin{equation}\label{eq.fad}
F_{\rm CMB}(\vec{k}_1,\vec{k}_2,\vec{k}_3)\equiv\frac{2}{3}\frac{\langle\delta_{\gamma}(\vec{k}_1)\delta_{\gamma}(\vec{k}_2)\delta_{\gamma}(\vec{k}_3)\rangle'}{P_\gamma(k_1)P_\gamma(k_2)+P_\gamma(k_1)P_\gamma(k_3)+P_\gamma(k_2)P_\gamma(k_3)},
\end{equation}
where $P_\gamma(k) = \langle\delta_{\gamma}(\vec{k})\delta_{\gamma}(-\vec{k})\rangle'$ is the CMB power spectrum and $'$s on correlators denote the fact that momentum-conserving factors of $(2\pi)^3\delta(\vec{k}_1+\cdots+\vec{k}_n)$ are stripped off. The prefactor $2/3$ ensures that for the adiabatic perturbations, the standard definition of primordial NG~\cite{Akrami:2019izv} is reproduced.
In the above example of single-source reheating, $\delta_{\rm GW}$ follows the same adiabatic perturbation as $\delta_\gamma$, and the bi-spectrum from the GW observation is
\begin{align}\label{eq.fgw}
F_{\rm GW}(\vec{k}_1,\vec{k}_2,\vec{k}_3)=&\frac{10}{9}\frac{\langle\delta_{\rm GW}(\vec{k}_1)\delta_{\rm GW}(\vec{k}_2)\delta_{\rm GW}(\vec{k}_3)\rangle'}{P_{\rm GW}(k_1)P_{\rm GW}(k_2)+P_{\rm GW}(k_1)P_{\rm GW}(k_3)+P_{\rm GW}(k_2)P_{\rm GW}(k_3)}\nonumber
\\
=&F_{\rm CMB}(\vec{k}_1,\vec{k}_2,\vec{k}_3),
\end{align}
where the factor of $10/9$ ensures equality with $F_{\rm CMB}$ for adiabatic perturbations. To obtain an estimate of the observational sensitivity of future GW detectors, we first note that given an experiment, the cosmic-variance-limited precision of bispectrum measurement is determined by the number of independent modes the experiment can observe. Hence, a crude cosmic variance limited sensitivity $\Delta F_{\rm GW}$ from the GWB observation which can access multipoles up to $\ell_{\rm max}$, is (see e.g.~\cite{Lyth:2009zz})
\begin{equation}\label{eq.fgwsens}
\Delta F_{\rm GW}\sim \frac{1}{\delta_{\rm GW}}\times \frac{1}{\sqrt{\ell_{\rm max}(\ell_{\rm max}+1)}}\sim 10^4\ell_{\rm max}^{-1}\,.
\end{equation}
Recalling that the search of the CMB bi-spectrum has set an upper bound $|F_{\rm CMB}|\lsim 5-50$~\cite{Akrami:2019izv} depending on the shape of NG, Eq.~\eqref{eq.fgwsens} implies that even if we can measure the anisotropy $\delta\rho_{\rm GW}$, the bounds on NG from GWB will be rather weak compared to that from CMB unless $\ell_{\rm max}\gsim 10^3$. However, probing scales such small angular scales corresponding to $\ell_{\rm max}\gsim 10^3$ can be quite difficult even considering DECIGO and BBO, given the previously discussed benchmark points.

\section{Non-Gaussian gravitational waves with isocurvature perturbations}
The above conclusions can significantly change once we consider the universe to go through three different stages of reheating, as put forward in~\cite{Geller:2018mwu}. The inflaton $\phi$, whose fluctuations are mostly Gaussian, decays into a Hidden Sector (HS) of particles which is distinct from the BSM sector undergoing the PT, and thermally decoupled from it. We assume that the BSM sector (which includes dark matter (DM)) is reheated by a separate field $\sigma$ which is a spectator field to the inflationary dynamics. We take both the inflationary expansion and primordial perturbations to be dominated by the inflaton $\phi$, and the $\sigma$ to contribute to isocurvature perturbations. This will enable us to have significantly NG $\sigma$-fluctuations while ensuring the observational bounds are satisfied.

\begin{figure}
    \centering
    \includegraphics[width=10cm]{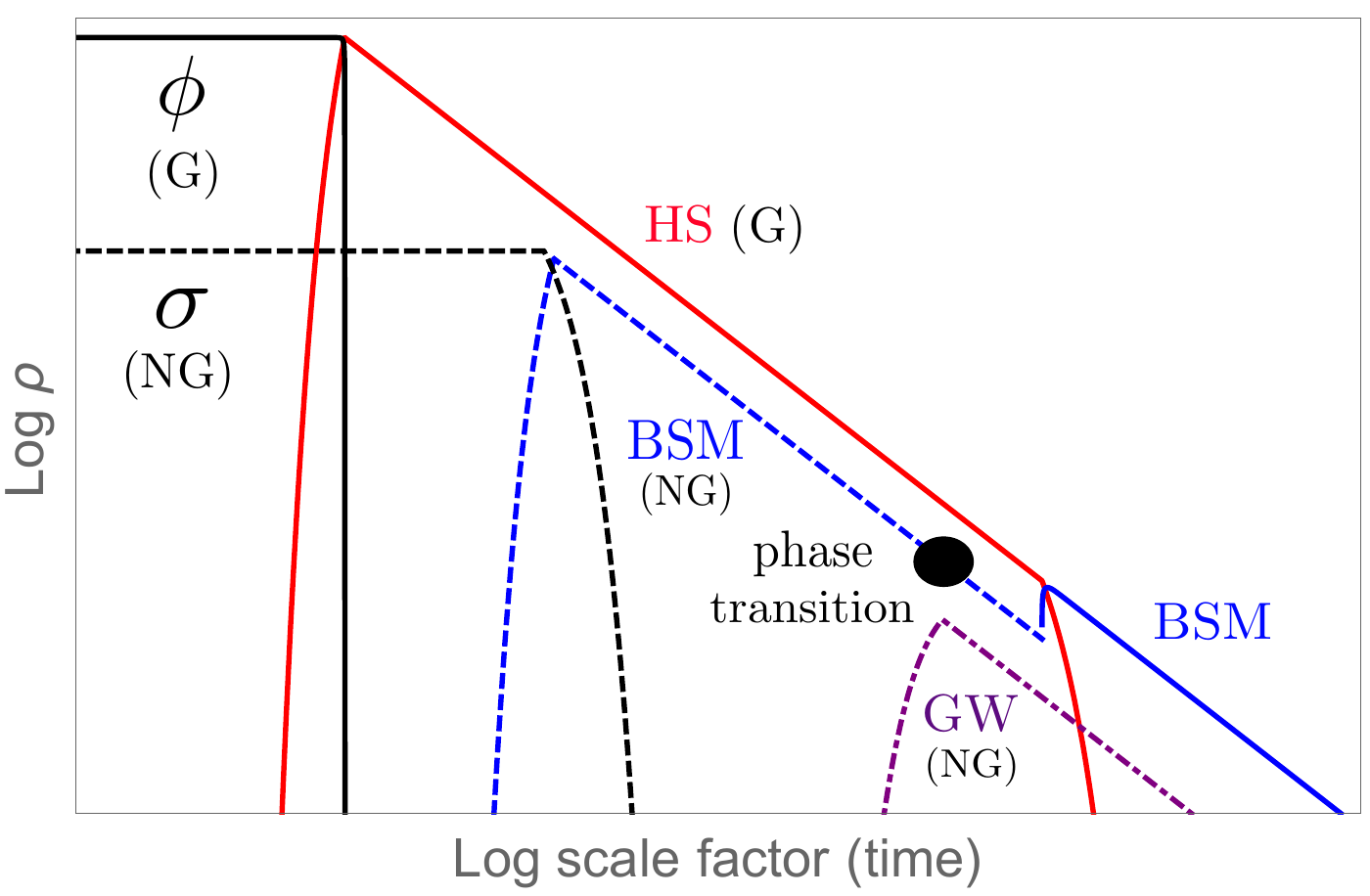}
    \caption{Cosmological history of the CMB and GWB production, along with their fluctuations. The inflaton field (solid black) carries a Gaussian (G) fluctuation and reheats a hidden sector (HS) (red). A separate field $\sigma$ (dashed black) carries a non-Gaussian (NG) fluctuation and reheats the BSM sector (blue) where the PT occurs. Following the PT, the HS decay into the BSM sector which eventually gives rise to the CMB. Depending on the NG of $\sigma$ and the size of $f_{\rm BSM}$, the fractional energy density of the BSM sector, the GW signal (purple) can have a large NG while CMB is almost Gaussian. The plot is adapted and modified from~\cite{Geller:2018mwu}.}
    \label{fig:reheating}
\end{figure}

After the PT, the HS also decays into the BSM sector and thermalizes with it. In this way, even though both reheating processes ultimately populate the BSM sector, if $\sigma$ reheats the universe to well above the PT temperature, and the HS decay happens at a temperature below the PT, GW anisotropies will receive contributions from the fluctuations of $\sigma$. Since GWs freely stream after getting generated, the reheating mediated by the HS decays does not affect the NG nature of the GW signal. In particular, any NG of $\sigma-$fluctuations will get imprinted and preserved in NG of GW. We illustrate the history of the two reheating processes in Fig.~\ref{fig:reheating}.

We label the energy densities in 
the two sectors (prior to HS decay) as $\rho_{\rm BSM}$ and $\rho_{\rm HS}$, with the associated energy fractions, 
\begin{equation}
f_{i}=\frac{\rho_{i}}{\rho_{\rm BSM}+\rho_{\rm HS}}\,,\quad i={\rm BSM,\,HS}\,.
\end{equation}
%Since HS was decoupled from the Visible Sector (VS) that goes through the PT, we label the energy density associated to $\phi$ by the HS, and the energy associated to $\sigma$ by the VS. We define the energy fraction 
We take $f_{\rm BSM}\ll f_{\rm HS}$ to enforce that the NG BSM sector contains a subdominant energy fraction. 

%We will ignore anisotropic stress of free-streaming fluids.\SK{Will find a better place for this sentence.}
To compute the NG in the present scenario we will use the gauge invariant comoving curvature perturbation on uniform-density hypersurfaces $\zeta$, to characterize cosmological perturbations~\cite{Malik:2008im}. In terms of the Newtonian gauge quantities \Eq{eq.new}, $\zeta$ is given by,
\begin{eqnarray}
     \zeta =-\Psi -H \frac{\delta\rho}{\dot{\rho}}, 
\end{eqnarray}
where $\rho$ is the total energy density and the dot denotes a derivative with respect to physical time $t$.
Furthermore, we define isocurvature perturbation of GW with respect to photons $\gamma$ in analogy with neutrino isocurvature perturbation, as
\begin{eqnarray}
    S_{\rm GW} \equiv 3(\zeta_{\rm GW}-\zeta_\gamma)\,,
\end{eqnarray}
where the total photon perturbation is
\begin{eqnarray}
\quad \zeta_\gamma\equiv f_{\rm BSM}  \zeta_{\gamma_{\rm BSM}}+f_{\rm HS}\zeta_{\gamma_{\rm HS}}\,.
\end{eqnarray}
In the above, we have denoted $\gamma_{\rm HS(\rm BSM)}$ to be the photons coming from the HS~(BSM) respectively, and
\begin{align}
    \zeta_i=-\Psi -H \frac{\delta\rho_i}{\dot{\rho}_i}, 
\end{align}
denotes the gauge invariant fluctuation of a single fluid $i$.
Since both $\sigma$ and $\phi$ ultimately reheat the BSM sector at high temperature where the SM and DM components are assumed to be still in equilibrium, there is no baryon, DM, or neutrino isocurvature perturbations from the reheating scenario. However, since GW are decoupled from the rest of the fluids after their  production, but before HS decay, it does carry isocurvature NG from the difference between the $\delta\sigma$ and $\delta\phi$ perturbations.

We assume the universe was radiation dominated (RD) when the PT takes place. After getting generated, GW climb out of the gravitational potential well created by fluctuations in $\phi$ and propagate to us. Thus, large-scale GW energy density fluctuations can be written in terms of primordial perturbations as (for the derivation, see the Appendix~\ref{sec.app}),
\begin{align}\label{eq.delgwo}
     \delgw = \delgwp+4\Phi_{\rm RD} = - \frac{4}{3}\zetarad + \frac{4}{3}(1-f_{\rm GW})\sgw,
\end{align}
where $\Phi_{\rm RD}$, $\zetarad$ are respectively the Newtonian potential and the primordial adiabatic curvature perturbation during RD. For later convenience, we rewrite the above in terms of fluctuations of $\phi$, $\zeta_\phi$ and of $\sigma$, $\zeta_\sigma$,
\begin{align}\label{eq.delgwoquant}
\delgw = -\frac{4}{3}\zeta_\phi+\frac{4}{3}S_\sigma\left(1-\frac{4}{3}f_{\rm BSM}\right),     
\end{align}
where the isocurvature fluctuations of $\sigma$ is defined as, $S_\sigma\equiv 3(\zeta_\sigma-\zeta_\phi)$.

On the other hand, large-scale CMB energy density fluctuations are given by (for a derivation see the Appendix~\ref{sec.app}),
\begin{align}\label{eq.delggo}
     \delgg = \delggp+4\Phi_{\rm MD} =  - \frac{4}{5}\zetarad +\frac{4}{15}f_{\rm GW} \sgw,
\end{align}
where we have approximated the Universe to be almost matter dominated (MD) during
CMB decoupling. This in terms of the field fluctuations read as,
\begin{align}\label{eq.delggoquant}
\delgg = -\frac{4}{5}\zeta_\phi-\frac{4}{15}f_{\rm BSM}S_\sigma.     
\end{align}
%In the above, we have ignored the contribution of GW since $f_{\rm GW} \equiv \rho_{\rm GW}/\rho_{\rm tot}\ll f_d$, as can be seen from \Eq{eq:rhoGW}, and $S_d = \sgw$. \YT{I didn't change the discussion around Eq.~(15)(16)}
%Let us first estimate the isocurvature contribution to the CMB power spectrum (suppressing the momentum argument on fluctuations)
%\begin{eqnarray}
%    P_\gamma = \left(\frac{4}{5}\right)^2\langle\zetarad^2\rangle \left(1+\left(\frac{f_d}{3}\right)^2\frac{\langle S_d^2\rangle}{\langle\zetarad^2\rangle}\right).
%\end{eqnarray}
Since the $\Delta N_{\rm eff}\lsim 0.4\,(2\sigma)$ constraint on the effective number of new relativistic degrees of freedom inferred 
from the CMB and Big-bang nucleosynthesis (BBN) analyses~\cite{Cyburt:2015mya,Aghanim:2018eyx} requires
\begin{eqnarray}\label{eq.fd}
     \rho_{\rm GW}\lsim 0.1\rho_{\gamma}\,,
\end{eqnarray}
this implies, together with Eq.~(\ref{eq:rhoGW}) with $\alpha\approx f_{\rm BSM}$, a weak upper bound on $f_{\rm BSM}\lesssim 1$.

%Here we have considered the case where the adiabatic and isocurvature perturbations are uncorrelated. %To our knowledge, there is no isocurvature bounds on CMB if isocurvature is sourced by DR \YT{there're still DR bounds under specific model building assumptions}. 
%\YTc{We can translate the neutrino isocurvature bound $S_\nu/\zetarad\lsim 0.1$,} into that of $S_d$ after weighting by the ratio of DR abundance and neutrino abundance (during RD) $f_d/f_\nu$ to get, 
%\begin{eqnarray}
%\frac{f_d}{f_\nu}\frac{S_d}{\zetarad}\lsim 0.1.   
%\end{eqnarray}
%Since $\Delta N_{\rm eff}$ bound already forces $f_d \lesssim 0.1$ as in \Eq{eq.fd} and $f_\nu\sim 1$, the above isocurvature bound can be satisfied with the same restriction on $f_d$, and $S_{d}\sim \zetarad$.

Using \Eqs{eq.delgwoquant}{eq.delggoquant}, we can now estimate the large-scale NG in CMB and GW by using \Eqs{eq.fad}{eq.fgw} respectively. To satisfy the NG bounds measured using CMB, from now on we will assume inflaton perturbations are mostly Gaussian. The CMB NG is determined by \Eqs{eq.fad}{eq.delggoquant} and in the limit $\zeta_\phi \gtrsim f_{\rm VS} S_\sigma$ they imply,
\begin{align}\label{eq:Fad}
F_{\rm CMB} \approx -\frac{5}{6}\left(\frac{f_{\rm BSM}}{3}\right)^3 \left(\frac{P_S(k_1)P_S(k_2)+\text{perms.}}{P_\phi(k_1)P_\phi(k_2)+\text{perms.}}\right)F_{S_\sigma},
\end{align}
where $P_\phi(k)=\langle\zeta_\phi(\vec{k})\zeta_\phi(-\vec{k})\rangle^\prime$ and $P_S(k)=\langle S_\sigma(\vec{k})S_\sigma(-\vec{k})\rangle^\prime$ are the inflaton and curvaton power spectrum respectively, and ``perms.'' include the permutations $(k_1,k_2)\rightarrow(k_1,k_3)$ and $(k_1,k_2)\rightarrow(k_2,k_3)$. The quantity $F_{S_\sigma}=\langle S_\sigma^3\rangle/(\langle S_\sigma^2\rangle^2+\rm perms.)$, schematically defined in analogy with \Eq{eq.fad} with $S_\sigma$ replacing $\delta_\gamma$ (but without the prefactor), characterizes isocurvature NG. We see that even if isocurvature perturbations are highly non-Gaussian with $F_{S_\sigma}\sim 1/S_\sigma \sim 10^4$, for $f_{\rm BSM}\lsim 0.1$ consistent with the $\Delta N_{\text{eff}}$ constraint~\Eq{eq.fd}, and $P_S\sim P_\phi$, we get $F_{\rm CMB}\lsim 10$, satisfying the NG bounds on CMB. 

However, this $f_{\rm BSM}^3$ suppression does not appear in GW NG. In particular,
\Eqs{eq.fgw}{eq.delgwo} imply for small $f_{\rm BSM}$,
\begin{align}\label{eq.fgw2}
\fgw = \frac{5\left(P_S(k_1)P_S(k_2)+\text{perms.}\right)F_{S_\sigma}}{6\left(P_\phi(k_1)+P_S(k_1)\right)\left(P_\phi(k_2)+P_S(k_2)\right)+\text{perms.}} ,
\end{align}
For a simple benchmark in which the two sectors have comparable primordial power spectra, $P_S\sim P_\phi$, we see
$\fgw\sim F_{S_\sigma}\sim 10^4$ then satisfies all the observational constraints. 

\begin{figure}
\includegraphics[width=12cm]{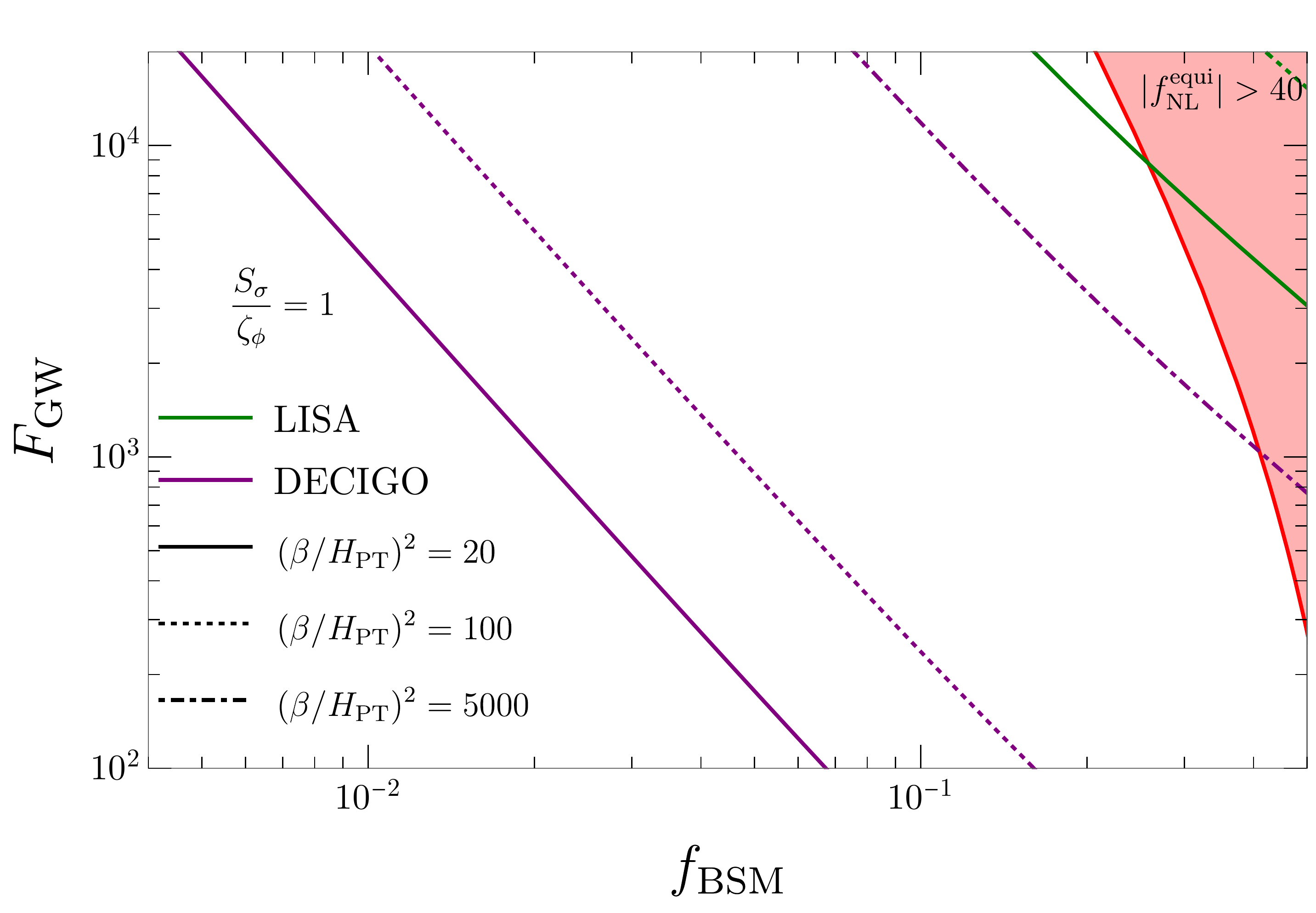}
\caption{The projected sensitivity of the Gravitational Wave Background NG, $F_{\rm GW}$ as a function of energy fraction $f_{\rm BSM}$ of the BSM sector undergoing PT for the case of equilateral NG. Depending on different assumptions of the PT parameter $(\beta/H_{\rm PT})$, LISA and DECIGO would be able to probe regions above the green and purple lines respectively. The reach of BBO, which we do not show explicitly for clarity, is expected to be slightly better than that of DECIGO. The region labeled $|f_{\rm NL}^{\rm equi}|>40$ is ruled out based on the constraints on equilateral NG from Planck~\cite{Akrami:2019izv}.} %\YT{define $f_{\rm NL}^{\rm GW}$, change one plot into $S_{\sigma}/\zeta_{\phi}<1$, change the range of $f^{GW}_{\rm NL}$ in the plot}}
\label{fig:fnl}
\end{figure}

As discussed in \Eq{eq.fgwsens}, $\fgw\sim 10^3-10^4$ can be detected with $\ell_{\rm max}\sim \mathcal{O}(10)$. Such values of $\ell_{\rm max}$, can be accessed in LISA for the benchmark point discussed earlier (again assuming suitable astrophysical foreground subtraction is performed). Hence, although crude, LISA itself might give us one of the first bounds, or a detection, of non-Gaussianity from GW. As discussed above, with DECIGO and BBO, one can access $\ell_{\rm max}\lsim \mathcal{O}(100)$. Correspondingly one can get to $\fgw\sim \mathcal{O}(100)$. Note, probing such a level of HS NG using CMB or LSS requires a $F_{\rm CMB} \sim 10^{-1}$ for $f_{\rm BSM}\sim 0.1$---challenging for LSS (given gravitationally induced non-linearities) and impossible for CMB for the equilateral shape.
%while difficult to access in LISA, can be accessed relatively easily with BBO as discussed before --- giving us a unique probe non-Gaussian interactions of HS, which would otherwise be invisible via CMB. 

Moving away from the above benchmark parameter choices, in Fig.~\ref{fig:fnl} we show projections of the LISA and DECIGO sensitivities on the $F_{\rm GW}$ measurements. We focus on the case of equilateral NG~\cite{Creminelli:2003iq,Babich:2004gb} for which GW can probe primordial NG in a powerful way, complementary to the CMB and LSS measurements. The limit of the $F_{\rm GW}$ measurement depends on the strength of the PT, the energy ratio $f_{\rm BSM}$, and the isocurvature perturbation $S_{\sigma}$. The NG signal is visible when both $\delta\Omega_{\rm GW}$ is higher than the detector sensitivity and $F_{\rm GW}\gsim\Delta F_{GW}$ in Eq.~(\ref{eq.fgwsens}), given good enough angular resolution.
The larger $f_{\rm BSM}$ region is excluded by a comparison of Eqs.~(\ref{eq.delom0}) and (\ref{eq:Fad}) to the the current bounds on $F_{\rm CMB}$ from the Planck (red) measurements. For equilateral NG, future LSS constraints from galaxy clustering (see e.g.~\cite{2012MNRAS.422.2854G}) are expected to be similar to the Planck constraints, and we do not show it explicitly. The $\Delta N_{\rm eff}$ constraints from the future CMB-S4 measurement~\cite{Abazajian:2016yjj} are too weak to be visible in Fig.~\ref{fig:fnl}. As we can see, a large non-Gaussian signal can exist in the GWB even with the current constraint, and the corresponding $\delta\rho_{\rm GW}$ can be within the reach of DECIGO and BBO. When the PT is strong source of GW, say with $(\beta/H)^2\sim 20$, it might even be possible for LISA to probe an interesting and novel part of the parameter space.
%and LSS measurements to see a large NG signal while satisfying the existing constraint. %[comment on different $S_{\sigma}/\zeta_{\phi}$}]
Lastly, in Fig.~\ref{fig:fnl} we have focused on the case $S_\sigma=\zeta_\phi$ just for simplicity, and we have checked that similar conclusions are obtained for other choices of $S_\sigma$.

\section{A non-Gaussian hidden sector model}
We now discuss a simple HS model that can give rise to large NG in GW.
We will assume that the spectator scalar field $\sigma$ is light during inflation so that it experiences significant quantum fluctuations, $m_\sigma^2\ll H_{\rm inf}^2$, where $H_{\rm inf}$ is the Hubble scale during inflation. We also assume the energy density in $\sigma$ both during and after inflation is subdominant. After the end of inflation, $\sigma$ dilutes like matter and eventually reheats into the BSM sector that later undergoes the PT. The isocurvature component of the BSM particles, before the HS decay into BSM, is then given by (see e.g.~\cite{Gordon:2000hv,Langlois:2011zz,Fonseca:2012cj}),
\begin{eqnarray}\label{eq.sd}
     S_{\sigma} = \frac{2\delta\sigma}{\sigma_0},
\end{eqnarray}
where $\sigma_0$ is the originally misaligned homogeneous VEV of the $\sigma$ field and $\delta\sigma$ denotes its fluctuation. For $S_\sigma\sim \zeta_\phi$ we need, $H_{\rm inf}/\sigma_0\sim 10^{-4}$. Around a temperature $T\sim \rm TeV$, the BSM sector undergoes a strong first-order PT and release GWs. At the end of the PT, the BSM sector consists of the remnant BSM plasma and the generated GW, and the only species eventually contributing to $\Delta N_{\rm eff}$ is the GW.

In the above set-up, any interaction involving $\sigma$ will induce NG in $\delta\sigma$, and eventually in GW. To characterize interactions of $\sigma$, while ensuring the radiative stability of its low mass, we will only consider a shift-symmetric derivative interaction,
\begin{align}\label{eq.sigmalag}
\mathcal{L}_\sigma = -\frac{1}{2}(\partial_\mu\sigma)^2-\frac{1}{2}m_\sigma^2\sigma^2+\frac{1}{\Lambda_\sigma^4}(\partial_\mu\sigma)^2(\partial_\nu\sigma)^2+\cdots.
\end{align}
At the classical level, we can get the equation of motion for the homogeneous background $\sigma_0$, which has a slow-roll approximation~\cite{Creminelli:2003iq}, $\ddot{\sigma}_0\ll H\dot{\sigma}_0$, under which
\begin{align}
\dot{\sigma}_0\left(1+4\frac{\dot{\sigma}_0^2}{\Lambda_\sigma^4}\right) \approx -\frac{m_\sigma^2}{3H_{\rm inf}}\sigma_0.
\end{align}
Hence to only have perturbative effects from the dimension-8 operator in \Eq{eq.sigmalag}, we will require $\dot{\sigma}_0^2/\Lambda_\sigma^4\lsim 0.1$.\footnote{Note, a choice of $\dot{\sigma}_0^2/\Lambda_\sigma^4\sim 0.1$ requires $H\lsim\Lambda_\sigma\ll \sigma_0$. Controlled effective field theory examples of such scenarios can be constructed using bi-axion alignment models~\cite{Kim:2004rp}, as explored in some detail in~\cite{Kumar:2019ebj}.} Importantly, the Lagrangian in  \Eq{eq.sigmalag} also contains a non-trivial cubic interaction term
\begin{align}\label{eq.cubicint}
\mathcal{L}_\sigma \supset 4 \frac{\dot{\sigma}_0}{\Lambda_\sigma^4}\dot{\delta\sigma}(\partial\delta\sigma)^2,
\end{align}
given which, the three-point function of $\delta\sigma$ can be calculated in the equilateral limit with $k_1=k_2=k_3=k$ as~\cite{Creminelli:2003iq},
\begin{align}
\langle\delta\sigma(\vec{k}_1)\delta\sigma(\vec{k}_2)\delta\sigma(\vec{k}_3)\rangle^\prime=-\frac{7}{3}\frac{\dot{\sigma}_0}{\Lambda_\sigma^4}\frac{H_{\rm inf}^5}{k^6}.
\end{align}
Using this, \Eq{eq.sd} and the definition of $F_{S_\sigma}$ given below~\Eq{eq:Fad} we can evaluate in the equilateral limit,
\begin{align}
|F_{S_\sigma}|=\frac{14}{9}\frac{H_{\rm inf}\sigma_0\dot{\sigma}_0}{\Lambda_\sigma^4}.
\end{align}
Therefore with the benchmark choice $\dot{\sigma}_0^2/\Lambda_\sigma^4\lsim 0.1$, $\dot{\sigma}_0\sim H_{\rm inf}^2$, $\sigma_0/H_{\rm inf}\sim 10^{4}$, we can have $\fgw\sim F_{S_\sigma} \lsim \text{few}\times 10^3$ which can be probed at LISA, and with much better precision at DECIGO and BBO.
 %As is clear from \Eq{eq:rhoGW}, only the remnant DR contributes dominantly to $\Delta N_{\rm eff}$.

\section{Conclusions} 
We have shown that large isocurvature non-Gaussianity (NG) can be hidden inside a stochastic GW background arising from a PT, and yet be invisible even to the near-future LSS or CMB measurements. If inflation and reheating involve multiple scalar fields, the reheating from some of the scalar fields can pick up large NG perturbations and produce NG GW background even when the rest of the cosmological fluids inherit predominantly Gaussian fluctuations. In this case, future surveys of the stochastic GW signal by LISA, DECIGO, and BBO may give us the first glimpse
of primordial NG quantum fluctuations in the early universe! 

%%%%%%%%%%%%%%%%%%%
~\\
\mysection{Acknowledgments}
We are grateful to Michael Geller, Anson Hook, Jonathan Kozaczuk, Arianna Renzini, Joseph Romano and especially Neil Cornish
for very useful discussions. This research was supported in part by the US-Israeli BSF Grant 2018236 and by the Maryland Center for Fundamental Physics. SK was also supported in part by the NSF grants PHY-1914731, PHY-1915314 and the U.S. DOE Contract DE-AC02-05CH11231. YT was also supported in part by the NSF grant PHY-2014165.
%\bibliography{ref}
%\onecolumngrid

\appendix
\setcounter{secnumdepth}{2}
\section{Derivations of large-scale CMB and GW anisotropy}\label{sec.app}
%\subsection{Generalities}
In this appendix we give a derivation of the large-scale anisotropy of CMB and GW in the context of the reheating scenario described in the main text. To this end, we denote the homogeneous metric using conformal time $\eta$ as,
\begin{equation}
ds^2 = a^2(\eta)(-d\eta^2+d\vec{x}^2),
\end{equation}
and the scalar metric fluctuations as~\cite{Malik:2008im},
\begin{align}
\delta g_{00}=-2a^2\varphi;\delta g_{0i}=a^2\partial_i B;\delta g_{ij}=2a^2(-\psi\delta_{ij}+\partial_i\partial_j E).
\end{align}
The relevant perturbed Einstein equations on superhorizon scales and in the absence of anisotropic stress are given by,
\begin{gather}
3\mathcal{H}(\mathcal{H}\varphi+\psi^\prime) = -4\pi G a^2\delta\rho,\nonumber\\
(E^\prime-B)^\prime+2\mathcal{H}(E^\prime-B)+\psi-\varphi = 0.
\end{gather}
Here ${}^\prime$s denote derivatives with respect to $\eta$.
The conformal Hubble rate and the perturbed energy density are respectively given by $\mathcal{H}$ and $\delta\rho$. As introduced in the main text, the curvature perturbation on uniform density hypersurface is given by the gauge invariant expression,
\begin{equation}
\zeta = -\psi-\mathcal{H}\frac{\delta\rho}{\rho'}.
\end{equation}

In the Newtonian gauge $B=E=0$, these equations simplify,
\begin{align}
3\mathcal{H}(\mathcal{H}\varphi+\psi') =& -4\pi G a^2\delta\rho,\label{eq.ein1}\\
\psi=& \phi\label{eq.ein2}.
\end{align}
Writing $\psi=\Psi$ and $\varphi=\Phi$ and using Eqs.~\eqref{eq.ein1} and \eqref{eq.ein2}, $\zeta$ can be written as,
\begin{align}\label{eq.zetaphi}
\zeta=-\Psi-\frac{2}{3(1+w)H}\left(H\Phi+\dot{\Psi}\right),
\end{align}
where $w$ is the effective equation of state.

We will be interested in obtaining expressions for large-scale modes. For this purpose, it will be useful to have the following relations between $\zeta$ and $\Phi(=\Psi)$ during both Radiation Domination (RD) and Matter Domination (MD):
\begin{eqnarray}
\zeta_{\rm RD} =& -\frac{3}{2}\Phi_{\rm RD},\label{eq.zetard}\\
\zeta_{\rm MD} =& -\frac{5}{3}\Phi_{\rm MD}\label{eq.zetamd}.
\end{eqnarray}
In the above, we have used the constancy of $\Phi$ on super-horizon scales in the absence of non-adiabatic pressure perturbation.
%\subsection{PT from a visible sector}
Now we apply the above results in the PT scenario described in the main text for which the PT takes place in a beyond Standard Model (BSM) sector reheated by $\sigma$. The hidden sector (HS), on the other hand, is reheated by inflaton, $\phi$. After the PT takes place and GW are released, the HS decays into BSM sector. %In this case, the only contribution to $\Delta\neff$ comes from gravitational waves.

\subsection{Large-scale CMB anisotropy}
The large-scale CMB anisotropies are given by, after accounting for the local redshift i.e., the SW effect,
\begin{align}
\tcmb = \frac{1}{4}\delta_\gamma^{\rm prim}+\Phi_{\rm MD}=\zeta_\gamma+2\Phi_{\rm MD}.
\end{align}
Here $\zeta_\gamma$ is the gauge invariant photon perturbation, generally given by $\zeta_\gamma=-\psi+\frac{1}{4}\delta_\gamma$.
% During matter domination the curvature perturbation is given by,
% \begin{align}
% \zeta=\zeta_{\rm MD}=-\Phi_{\rm MD}+\frac{1}{3}\delta_m.
% \end{align}
Using Eq.~\eqref{eq.zetamd} this gives,
\begin{align}
\tcmb=\zeta_\gamma-\frac{6}{5}\zeta_{\rm MD}.
\end{align}
Since in this scenario both the matter and neutrino number densities track the photon number density, the only isocurvature is in GW. Thus we have $\zeta_{\rm MD}=\zeta_\gamma$ implying
\begin{align}
\tcmb=-\frac{1}{5}\zeta_\gamma.
\end{align}

To compare with the Planck isocurvature bounds it is more convenient to write $\tcmb$ as a linear combination of curvature and isocurvature perturbations. Using the fact that after HS decay, we only have photons, neutrinos ($\nu$) and GW diluting as radiation,
\begin{align}
\zeta_{\rm RD}=&(1-f_\nu-f_{\rm GW})\zeta_\gamma+f_\nu\zeta_\nu+f_{\rm GW}\zeta_{\text{GW}}\nonumber\\ 
=& \zeta_\gamma+\frac{1}{3}f_{\rm GW} \sgw,
\end{align}
we get
\begin{align}
\tcmb=-\frac{1}{5}\zeta_{\rm RD}+\frac{1}{15}f_{\rm GW}\sgw.
\end{align}
In the above we have used the energy density fractions, $f_i = \rho_i/\rho_{\rm tot}$ for the three species $i=\gamma,\nu,\text{GW}$ with $\rho_{\rm tot}$ being the total energy density during RD. %We see for a benchmark choice $\sgw\sim\zetarad$, if we ensure $f_{\rm GW} \lesssim 0.1$, we can satisfy the Planck isocurvature bounds. 
To compute NG, on the other hand, it is more convenient if we express $\tcmb$ in terms of uncorrelated inflationary perturbations. For this purpose we write,
\begin{align}
\delta_{\gamma}\equiv4\tcmb=& -\frac{4}{5}\left(\zeta_{\gamma_{\rm HS}}+f_{\rm BSM}(\zeta_{\gamma_{\rm BSM}}-\zeta_{\gamma_{\rm HS}})\right)\nonumber\\
=& -\frac{4}{5}\zeta_{\phi}-\frac{4}{15}f_{\rm BSM}S_{\sigma},
\end{align}
To obtain the last relation we have used $\zeta_{\rm HS}=\zeta_\phi$ and $\zeta_{\rm BSM}=\zeta_\sigma$ since the HS and the BSM sector respectively inherit the fluctuations of $\phi$ and $\sigma$ fields. We have also used the standard definition $S_{\sigma}\equiv3(\zeta_\sigma-\zeta_\phi)$.

%In the regime where $\sigma$ is a spectator field, $\zeta_{\gamma_2}=\zeta_\phi$ and $S_{\gamma_1}=2\delta\sigma/\sigma_0$. This is because during inflation $\sigma$ does not appreciably contribute to energy density, and the spacetime curvature is mostly determined by the inflaton perturbation $\zeta_\phi$. Hence during HS reheating, the $\gamma_2$ photon perturbations pick up this same $\zeta_\phi$ with the assumption that $\sigma$ is still contributes subdominantly to energy density. Finally, the isocurvature perturbations in $\sigma$, and hence in GW, is determined by $\delta\sigma$. Therefore,
% \begin{align}
% F_{\rm CMB}\sim& \frac{5}{4}\left(\frac{f_{\gamma_1}}{3}\right)^3\frac{\langle S_{\gamma_1}^3\rangle}{\langle\zeta_{\gamma_2}^2\rangle^2}\nonumber\\
% \sim & \frac{5}{4}\left(\frac{f_{\gamma_1}}{3}\right)^3 F_{S} \frac{\langle S_{\gamma_1}^2\rangle^2}{\langle\zeta_{\gamma_2}^2\rangle^2}.
% \end{align}

\subsection{Large-scale GW anisotropy}
Now we repeat the calculation for GW perturbations. Taking into account the redshift and the fact the GWs were generated during RD we can again write,
\begin{align}
\tgw = \frac{1}{4}\delta_{\rm GW}^{\rm prim}+\Phi_{\rm RD}=\zeta_{\rm GW}-\frac{4}{3}\zetarad.
\end{align}
Using the fact that $\zetarad=\zeta_\gamma+\frac{1}{3}f_{\rm GW}\sgw$,
\begin{align}
\tgw=-\frac{1}{3}\zetarad+\frac{1}{3}(1-f_{\rm GW})\sgw.
\end{align}
As before, it is convenient to write the above in terms of uncorrelated perturbations,
\begin{align}
\delta_{\rm GW}\equiv 4\tgw=&-\frac{4}{3}\zetarad+\frac{4}{3}(1-f_{\rm GW})\sgw\nonumber\\
\approx& -\frac{4}{3}\zeta_\phi+\frac{4}{3}S_\sigma \left(1-\frac{4}{3}f_{\rm BSM}\right),
% =&-\frac{4}{3}\zeta_\gamma+\frac{4}{3}\left(1-\frac{4}{3}f_{\rm GW}\right)\sgw\nonumber\\
% =&-\frac{4}{3}\zeta_{\gamma_2}+\frac{4}{3}\left(f_{\gamma_2}\left(1-\frac{4}{3}f_{\rm GW}\right)-\frac{1}{3}f_{\gamma_1}\right)S_{\gamma_1}.
\end{align}
where to get to the last expression we have used, $f_{\rm GW}\ll 1$.
% Thus the NG contribution is,
% \begin{align}
% \fgw\sim\frac{3}{4}\frac{\langle\sgw^3\rangle}{\left(\langle\zeta_{\gamma_2}^2\rangle+\langle\sgw^2\rangle\right)^2}.
% \end{align}
% This does not have the $f_{\gamma_1}^3$ suppression as expected, and hence, GW NG can be much bigger while ensuring small backreacted CMB NG.

\bibliography{AGW.bib}

\end{document}